\documentstyle[12pt,aps]{revtex}
\begin{document}
\author{Yishi Duan$^1$, Libin Fu$^{2,1,}$\thanks{%
Corresponding author. E-mail: lbfu@263.net} and Hong Zhang}
\address{$^1$Physics Department, Lanzhou University, Lanzhou,Gansu,730000, China \\
$^2$Institute of Applied Physics and Computational Mathematics \\
P.O.Box 8009(26), Beijing 100088, China}
\title{Topological quantum mechanics and the first Chern class }
\date{June 18, 1999}
\maketitle

\begin{abstract}
\begin{center}
{\bf Abstract}
\end{center}

Topological properties of quantum system is directly associated with the
wave function. Based on the decomposition theory of gauge potential, a new
comprehension of topological quantum mechanics is discussed. One shows that
a topological invariant, the first Chern class, is inherent in the
Schr\"odinger system, which is only associated with the Hopf index and
Brouwer degree of the wave function. This relationship between the first
Chern class and the wave function is the topological source of many
topological effects in quantum system.
\end{abstract}

\newpage
\section{Introduction}

Topology now becomes absolutely necessary in physics$^{1-3}$. The $\phi $%
-mapping theory and the gauge potential decomposition theory$^{4-7}$ are
found to significant in exhibiting the topological structure of physics
system and have been used to study topological current of magnetic monopole$%
^{4,8}$, topological string theory$^6$, topological structure of
Gauss-Bonnet-Chern theorem$^9$, topological structure of the SU(2) Chern
density$^{10}$, topological characteristics of dislocations and
disclinations continuum$^{11,12}$, topological structure of the defects of
space-time in early universe as well as its topological bifurcation$^{13,14}$%
.

Topological properties of quantum system should be directly associated with
the wave function. Recently, using the $\phi $-mapping theory, the
topological structure of the London equation in superconductor has been
studied$^{15}$. It is showed that the topological structure of London
equation is characterized by topological index of wave function.

In this paper, based on $\phi $-mapping theory and gauge potential
decomposition theory, we reveal the inner relation between the topological
property of Schr\"odinger system and the intrinsic properties of its wave
function. For the first time, we point out that a topological invariant, the
first Chern class, is inherent in the Schr\"odinger system, which is only
associated with the wave function, without using any particular models or
hypotheses. One can find that this relationship between the first Chern
class and the wave function is the topological source of the inner structure
of London equations in superconductor$^{15}$.

\section{Decomposition theory of $U$(1) gauge potential and the first Chern
class}

Considering a complex line bundle ${\bf R}^3{\bf \times C,}$ as one knows,
the $U(1)$ gauge potential $A=A_idx^i$ is a connection on this bundle. A
section of the line bundle gives a complex valued functions $\psi ,$ and the
convariant derivative on the line bundle is defined as%
$$
D\psi =d\psi -iA\psi , 
$$
and its complex conjugate 
$$
D\psi ^{*}=d\psi ^{*}+iA\psi ^{*}. 
$$
From these formula above, one can obtain 
\begin{equation}
\label{ab}A=-\frac i2\frac{d\psi ^{*}\psi -d\psi \psi ^{*}}{\psi \psi ^{*}}+%
\frac i2\frac{D\psi ^{*}\psi -D\psi \psi ^{*}}{\psi \psi ^{*}}. 
\end{equation}
The main feature of the decomposition theory of the gauge potential is that
the gauge potential $A$ can be generally decomposed as 
\begin{equation}
\label{ab0}A=a+b, 
\end{equation}
where $a$ is required to satisfy the gauge transformation rule and $b$
satisfies the vector convariant transformation, i.e., 
\begin{equation}
\label{ag}a^{\prime }=a+d\alpha 
\end{equation}
and 
\begin{equation}
\label{bg}b^{\prime }=b 
\end{equation}
under $U(1)$ transformation $\psi ^{\prime }=e^{i\alpha }\psi .$ One can
show that the gauge potential $A$ are rigorously satisfies the gauge
transformation%
$$
A^{\prime }=A+d\alpha . 
$$
Comparing (\ref{ab0}) with (\ref{ab}), we can obtain a decomposition
expression of $U(1)$ gauge potential by defining 
\begin{equation}
\label{a}a=-\frac i2\frac{d\psi ^{*}\psi -d\psi \psi ^{*}}{\psi \psi ^{*}}, 
\end{equation}
and 
\begin{equation}
\label{b}b=\frac i2\frac{D\psi ^{*}\psi -D\psi \psi ^{*}}{\psi \psi ^{*}}. 
\end{equation}
One can easily prove that this decomposition satisfies the transformation
rules (\ref{ag}) and (\ref{bg}).

We know that the complex valued function $\psi $ can be denoted as 
$$
\psi =\phi ^1+i\phi ^2, 
$$
in which $\phi ^1$ and $\phi ^2$ are real valued function and can be
regarded as two components of a two-dimensional vector field ${\bf \phi }%
=(\phi ^1,\phi ^2)$ on ${\bf R}^3.$ The unit vector field is defined as 
\begin{equation}
\label{unit}n^a=\frac{\phi ^a}{||\phi ||},\quad ||\phi ||=(\phi ^a\phi ^a)^{%
\frac 12},\quad a=1,2, 
\end{equation}
satisfying%
$$
n^an^a=1. 
$$
From (\ref{a}) and (\ref{unit}), it can be seen that%
$$
a=-\epsilon _{ab}dn^an^b. 
$$

We know that the characteristic class is the fundamental topological
property, and it is independent of the gauge potential$^{1,16}$. So, to
discuss the Chern class, we can take $A$ as 
\begin{equation}
\label{aa}A=-\epsilon _{ab}dn^an^b. 
\end{equation}
One can regard it as a special gauge. Then the field strength (the
curvature) $F$ can be expressed as 
\begin{equation}
\label{f}F=dA=\epsilon _{ab}dn^a\wedge dn^b. 
\end{equation}
Using (\ref{unit}) and 
$$
dn^a=\frac{d\phi ^a}{||\phi ||}-\frac{\phi ^ad(||\phi ||)}{||\phi ||^2}, 
$$
$$
\frac \partial {\partial \phi ^a}\ln ||\phi ||=\frac{\phi ^a}{||\phi ||^2} 
$$
$F$ changes into%
$$
F=\epsilon _{ab}\frac \partial {\partial \phi ^c}\frac \partial {\partial
\phi ^a}\ln ||\phi ||d\phi ^c\wedge d\phi ^b. 
$$
By making use of the Laplacian relation in $\phi $ space:%
$$
\frac \partial {\partial \phi ^a}\frac \partial {\partial \phi ^a}\ln ||\phi
||=2\pi \delta ^2({\bf \phi }), 
$$
we obtain 
\begin{equation}
\label{fdelt}F=2\pi \delta ^2({\bf \phi })\epsilon ^{ab}d\phi ^a\wedge d\phi
^b, 
\end{equation}

One finds that $F$ does not vanish only at the zero points of ${\bf \phi ,}$
i.e. 
\begin{equation}
\label{so}{\bf \phi (x)=}0. 
\end{equation}
The solution of Eqs. (\ref{so}) are generally expressed as 
\begin{equation}
\label{so1}{\bf x=x}_i(v),\quad i=1,\cdots m, 
\end{equation}
which represent $m$ zero lines $L_i$ $(i=1,\cdots ,m)$ with $v$ as intrinsic
coordinates. There exists a two-dimensional manifold $\Sigma $ which
normally intersects $L_i$ at the point ${\bf x}_i$ and ${\bf u=(}u_1,u_2)$
are the intrinsic coordinates on $\Sigma .$ In the $\delta $-function theory$%
^{17,18}$, one can prove that 
\begin{equation}
\label{de}\delta ^2({\bf \phi })=\sum_{i=1}^m\beta _i\eta _i\int_{L_i}\frac{%
\delta ^3({\bf x}-{\bf x}_i(v))}{D(\phi /u)_\Sigma }dv, 
\end{equation}
where 
\begin{equation}
\label{ja}D(\phi /u)_\Sigma =\frac 12\epsilon ^{ij}\epsilon _{ab}\frac{%
\partial \phi ^a}{\partial u^i}\frac{\partial \phi ^b}{\partial u^j}. 
\end{equation}
The positive $\beta _i$ is the Hopf index of $\phi $-mapping and $\beta _i$
is the Brouwer degree$^{19}$: 
$$
\eta _i=sgnD(\frac \phi u)_\Sigma =\pm 1 
$$
The meaning of the Hopf index $\beta _i$ is that the vector field function $%
{\bf \phi }$ covers the corresponding region $\beta _i$ times while ${\bf x}$
covers the region neighborhood of zero ${\bf z}_i$ once.

The integration of $F$ on $\Sigma $%
\begin{equation}
\label{c1}C_1=\frac 1{2\pi }\int_\Sigma F 
\end{equation}
is the first Chern class, an important topological invariant of the line
bundle. From (\ref{fdelt}) and (\ref{de}), we can obtain 
\begin{equation}
\label{c11}C_1=\sum_{i=1}^m\beta _i\eta _i. 
\end{equation}
From this result, we see that the first Chern class is the sum of the index
of zero points, and labeled by the Hopf index and Brouwer degree, or the
Winding number of ${\bf \phi .}$

\section{Topological invariant in quantum mechanics}

The topological property of quantum system should be directly associated
with the intrinsic properties of the wave function. Considering a
Schr\"odinger system:%
$$
i\hbar \frac \partial {\partial t}\psi =H\psi , 
$$
where $\psi $ is the wave function. The current density of this system is
given by%
$$
{\bf j(r,}t)=-\frac{i\hbar }{2m}(\psi ^{*}\nabla \psi -\psi \nabla \psi
^{*}). 
$$
For the purpose to study the topological property, we consider this system
with a fixed time. Hence the wave function $\psi $ can be regarded as a
section of a complex line bundle over ${\bf R}^3$ and can be denoted as 
\begin{equation}
\label{wa}\psi ({\bf r})=\phi ^1({\bf r})+i\phi ^2({\bf r}). 
\end{equation}
One defines a physical quantity ${\bf V}$ as 
\begin{equation}
\label{vvvv}{\bf V=-}\frac i2\frac{(\psi ^{*}\nabla \psi -\psi \nabla \psi
^{*})}{\psi \psi ^{*}}. 
\end{equation}
Under $U(1)$ transformation $\psi ^{\prime }=e^{i\alpha }\psi ,$ one can see
that ${\bf V}$ satisfies the gauge transformation%
$$
{\bf V}^{\prime }{\bf =V+}\nabla \alpha . 
$$
So, one should notice here that ${\bf V}$ is a composed $U(1)$ gauge
potential. One can prove that 
\begin{equation}
\label{vv}{\bf V=-\epsilon }_{ab}\nabla n^an^b=-{\bf \epsilon }_{ab}\partial
_in^an^b\vec e_i, 
\end{equation}
where $\vec e_i$ $(i=1,2,3)$ denote $(\hat x,\hat y,\hat z),$ and 
$$
\nabla \times {\bf V=\epsilon }^{ijk}\epsilon _{ab}\partial _jn^a\partial
_kn^b\vec e_i=2\pi J^i\vec e_i, 
$$
with 
\begin{equation}
\label{cc}J^i=\frac 1{2\pi }{\bf \epsilon }^{ijk}\epsilon _{ab}\partial
_jn^a\partial _kn^b, 
\end{equation}
which is a topological current. From the discussion in above section, we
have 
\begin{equation}
\label{vs}\nabla \times {\bf V=}2\pi \delta ^2({\bf \phi })\vec D(\frac \phi 
x), 
\end{equation}
where 
$$
\vec D(\frac \phi x)=\epsilon ^{ijk}\epsilon _{ab}\partial _j\phi \partial
_j\phi \vec e_i. 
$$
And then 
\begin{equation}
\label{vvv}\nabla \times {\bf V=}2\pi \sum_{i=1}^m\beta _i\eta
_i\int_{L_i}\delta ^3(\vec r-\vec r_i(v))\frac{\vec D(\frac \phi x)}{D(\phi
/u)_\Sigma }dv. 
\end{equation}
From$^7$, it is easy to see that%
$$
\frac{\vec D(\frac \phi x)}{D(\phi /u)_\Sigma }|_{\vec r_i(v)}=\frac{d\vec r%
_i}{dv}, 
$$
then the current (\ref{cc}) is turned to 
\begin{equation}
\label{vor}\vec J=\frac 1{2\pi }\nabla \times {\bf V=}\sum_{i=1}^m\beta
_i\eta _i\int_{L_i}\delta ^3(\vec r-\vec r_i)d\vec r_i. 
\end{equation}
It is obvious to see that the formula (\ref{vor}) represents a current of $m$
isolated vortices with the $i$-th vortex carries charge $2\pi \beta _i\eta
_i.$ And, one can prove that 
\begin{equation}
\label{c1q}Q=\int_\Sigma \vec J\cdot d\vec \sigma =\sum_{i=1}^m\beta _i\eta
_i. 
\end{equation}
Comparing this result with (\ref{c1}) and (\ref{c11}), we see that the total
topological charge of this system is equal to the first Chern number. So,
one finds that the Schr\"odinger system inherits a topological invariant,
the first Chern class, which is only associated with the intrinsic
properties of wave function, without using any particular models or
hypotheses. One can find that this relationship between the first Chern
class and the wave function is the source of many topological effects in
quantum system.

As an example, let us now consider a homogeneous superconductor in a
magnetic field which is weak compared with the critical field $B_{c_2}$ at
which the superconductivity is lost. The relation between the superconductor
current $\vec j_s$ and the condensate wave function $\psi $ is 
\begin{equation}
\label{tqj}{\bf j}_s=\frac{e\hbar }\mu |\psi |^2{\bf V-}\frac{2e^2}{\mu c}%
|\psi |^2{\bf A.} 
\end{equation}
Let the body we studied be in a state of thermodynamic equilibrium, so that
there is no normal current and ${\bf j}={\bf j}_s.$ We shall also use the
general Maxwell's equations: 
\begin{equation}
\label{max1}\nabla \times {\bf B=}\frac{4\pi }c{\bf j,} 
\end{equation}
\begin{equation}
\label{max2}\nabla \cdot {\bf B=}0. 
\end{equation}
To put them in the appropriate form, we first rewrite the relation (\ref{tqj}%
) between the superconductivity current density and ${\bf V}$ through (\ref
{max1}): 
\begin{equation}
\label{lon1}{\bf A}+\lambda ^2\nabla \times {\bf B}=\frac{\phi _0}{2\pi }%
{\bf V.} 
\end{equation}
For the London approximation corresponds to the assumption that $\lambda $
is constant, taking the curl of both sides of (\ref{lon1}) and noting that $%
\nabla \times {\bf A}={\bf B}$ and (\ref{max2}), we have 
\begin{equation}
\label{lon2}{\bf B}-\lambda ^2\nabla ^2{\bf B}=\frac{\phi _0}{2\pi }\nabla
\times {\bf V.} 
\end{equation}
Comparing this expression with (\ref{vor}), one obtain the topological
structure of London equation: 
\begin{equation}
\label{lon3}{\bf B}-\lambda ^2\nabla ^2{\bf B}=\phi _0\sum_{i=1}^m\beta
_i\eta _i\int_{L_i}d{\bf r}\delta ^3({\bf r}-{\bf r}_i). 
\end{equation}
It is obvious to see that the equation (\ref{lon3}) represents $m$ isolated
vortices of which the $i$th vortex carries flux $\beta _i\eta _i\phi _0.$ On
can conclude that includes vortex-antivortex pair$^{20,21}(\beta _1=\beta
_2, $ $\eta _1=1$ and $\eta _2=-1),$ vortex rings ($L_i$ is a ring),
multicharged vortices$^{22}$.

\section{Discussion}

Based on the $\phi $-mapping theory, using gauge potential decomposition
method, we reveal that the first Chern class is inherent in Schr\"odinger
system, which is only associated with the intrinsic properties of the wave
function. In fact, this topological property does not only relate to the
Schr\"odinger system, but also relates to the non-linear Schr\"odinger
system as well as the condensate Schr\"odinger system. We must point out
that this relationship naturally exists in $(2+1)$-dimensional quantum
system and is associated with the discussion of topological structure of
quantum Hall effect$^{23}$. We believe that this intrinsic topological
property is a fundamental property of quantum system and is the source of
many topological effects in quantum system.

\section*{Acknowledgments}

This work was supported by the National Natural Science Foundation of China.

\vskip 1cm

$^1$S. Nash and S. Sen, {\it Topology and Geometry for Physicists}
(Academic, London, 1983).

$^2$G. Morandi, {\it The Role of Topology in Classical and Quantum Physics},
Lecture Note in Physics Vol. {\bf M7} (Springer-Verlag, Berlin, 1992).

$^3$A. P. Balachandran, Found. Phys. {\bf 24}(4), 455 (1994).

$^4$Y. S. Duan and M. L. Ge, Sci. Sinica {\bf 11}, 1072 (1979).

$^5$Y. S. Duan, SLAC-PUB-3301/84.

$^6$Y. S. Duan and J.\ C. Liu, {\it Proceedings of Johs Hopkins Workshop 11}
(World Scientific, Singapore, 1988).

$^7$Y. S. Duan and X. H. Meng, J. Math. Phys. {\bf 34}(3), 4463 (1993).

$^8$G. H. Yang and Y. S. Duan, Inter. J. Theor. Phys. {\bf 37} (1998) 2435.

$^9$Y. S. Duan and X. H. Meng, J. Math. Phys. {\bf 34} (1993) 1149; Y. S.
Duan, S. Li and G. H. Yang, Nucl. Phys. {\bf B 514} (1998) 705.

$^{10}$Y. S. Duan and L. B. Fu, J. Math. Phys. {\bf 39 }(1998) 4343.

$^{11}$Y. S. Duan and S. L. Zhang, Int. J. Eng. Sci{\it .} {\bf 28,} 689
(1990); {\bf 29,} 153 (1991); {\bf 29}, 1593 (1991); {\bf 30,} 153 (1992).

$^{12}$Y. S. Duan and X. H. Meng, Int. J. Engng. Sci{\it .}{\bf \ 31, }1173
(1993).

$^{13}$Y. S. Duan, S. L. Zhang, and S. S. Feng, J. Math. Phys{\it .} {\bf 35}
(1994){\bf \ }4463; Y. S. Duan, G. H. Yang, and Y. Jiang, Int. J. Mod. Phys. 
{\bf A} {\bf 58} (1997) 513.

$^{14}$Y. S. Duan, G. H. Yang, and Y. Jiang, Gen. Rel. Grav. {\bf 29}, 715
(1997).{\bf \ }

$^{15}$Y. S. Duan, H. Zhang and S. Li, Phys. Rev. {\bf B} {\bf 58} (1998)
125.

$^{16}$M. W. Hirsch, {\it Differential Topology} (Springer verlag. New York
1976).

$^{17}$J. A. Schouten, Tensor Analysis for Physicists (Clarendon, Oxford,
1951).

$^{18}$G. Y. Yang, Ph.D. thesis, Lanzhou University, China (1997).

$^{19}$H. Hopf, Math. Ann. {\bf 96} 209 (1929).

$^{20}$B. A. Dubrovin, A. T. Fomenko and S. P. NoviKov, {\it Modern Geometry
and Applications} (Springer-Verlag, New York, 1984).

$^{21}$V. L. Berezinsky, Zh. \'Eksp. Tero. Fiz. {\bf 59}, 907 (1970) [Sov.
Phys. JEPT {\bf 32, 493 }(1971){\bf ].}

$^{22}$I. Aranson and V. Steinberg, Phys. Rev. {\bf B 53}, 75 (1996).

$^{23}$Y. S. Duan and S. Li, Phys. Lett. {\bf A 246}, 172 (1998).

\end{document}